\newcommand{\bef}{\begin{figure}}
\newcommand{\eef}{\end{figure}}
\newcommand{\bc}{\begin{center}}
\newcommand{\ec}{\end{center}}

\newcommand{\be}{\begin{equation}}
\newcommand{\ee}{\end{equation}}
\newcommand{\bea}{\begin{eqnarray}}
\newcommand{\eea}{\end{eqnarray}}

\documentclass[review]{elsarticle}

\usepackage{lineno,hyperref,amsmath,graphicx,amssymb}
\modulolinenumbers[5]

\journal{Chaos, Solitons \& Fractals}

\begin{document}

\begin{frontmatter}

\title{Towards predicting COVID-19 infection waves: A random-walk 
Monte Carlo simulation approach}


\author[utkal]{D.\,P.~Mahapatra}
\ead{dpm.iopb@gmail.com}

\author[uwc]{S.~Triambak}
\ead{striambak@uwc.ac.za}

\address[utkal]{Department of Physics, Utkal University, Vani Vihar, Bhubaneshwar 751004, India}
\address[uwc]{Department of Physics and Astronomy, University of the Western Cape, P/B X17, Bellville 7535, South Africa}

\begin{abstract}
Phenomenological and deterministic models are often used for the estimation of 
transmission parameters in an epidemic and for the prediction of its growth trajectory. 
Such analyses are usually based on single peak outbreak dynamics.   
In light of the present COVID-19 pandemic, there is a pressing need to better understand observed epidemic growth with multiple peak structures, preferably using first-principles methods. 
Along the lines of our previous work ~[Physica A {\bf 574}, 126014 (2021)], here we apply 2D random-walk Monte Carlo calculations to better understand COVID-19 spread through contact interactions. Lockdown scenarios and all other control interventions are imposed through mobility restrictions and a regulation of the infection rate within the stochastically interacting population. The susceptible, infected and recovered populations are tracked over time, with daily infection rates obtained without recourse to the solution of differential equations. 

The simulations were carried out for population densities corresponding to four countries, India, Serbia, South Africa and USA. In all cases our results capture the observed infection growth rates. More importantly, the simulation model is shown to predict secondary and tertiary waves of infections with reasonable accuracy. This predictive nature of multiple wave structures provides a simple and effective tool that may be useful in planning mitigation strategies during the present pandemic.

\end{abstract}


\end{frontmatter}


\section{Introduction}
\label{intro}
In the midst of the current COVID-19 pandemic, there is a continuing need to accurately model region-specific infection and mortality data, so that intervention methods and containment strategies can be planned accordingly. 
The simplest picture of epidemic growth is at most times provided by phenomenological models that are based on logistic growth~\cite{richards,Roosa,wu20,singer,shen,chowel,pwg}. However, real-time interventions that may affect the trajectory of the growth curve are not incorporated in such models.
%
%
%
%
More commonly, epidemiological modeling uses compartmentalized 
populations based on the SIR model~\cite{SIR} and its variants (see, for example~\cite{maier,axel,He}) that follow the time evolution of susceptible $(S)$, infected $(I)$ and recovered
$(R)$ populations, among others~\cite{sutra}. 
More often than not these deterministic models do not involve a stochastic formulation, 
which takes into account the random aspects of human mobility. Based on the early work by Bartlett~\cite{Bartlett}, such randomness can be incorporated through the equations~\cite{Fofana,Bailey_book,Noble}
\begin{align}
\frac{\partial S}{\partial t} = D_S \nabla^2 S -\beta S I\\
\frac{\partial I}{\partial t} = D_I \nabla^2 I + \beta S I -\mu I.
\end{align}
Here, $\nabla^2 = \left(\frac{\partial^2}{\partial x^2} + \frac{\partial^2}{\partial y^2}\right)$, and $S(x,y,t)$ and $I(x,y,t)$ are the spatial densities of the susceptible and infectious components of the population. The $D_{S,I}\nabla^2(S,I)$ `diffusion' terms represent the spatial movement of both susceptibles as well as infectives. The parameters $\beta$ and $\mu$ represent the infection and recovery rates respectively. Such formulations have also been extended to biased~\cite{Codling} and L\'evy~\cite{Raichlin} random walks. More recently, stochastic models have been developed using Markovian chains~\cite{ziqi,Gourieroux}, contact and community networks~\cite{network,Hoen}, Bayesian modeling~\cite{Dehning} etc. The effects of mobility restrictions have also been studied using Monte Carlo techniques~\cite{Sausa, st,turkey} that inherently include the diffusion terms mentioned above. 
In most studies, the analyses that are solely based on single epidemic growth curves do not adequately explain the multiple wave structures seen in global COVID-19 data. For example, it was recently proposed that a superposition of epidemic waves~\cite{Koltsova} could be used to describe COVID-19 growth curves. However this does not describe the observed multiple waves of infection for individual countries, that are well separated in time. 

The present work uses an uncorrelated random walk approach to study COVID-19 infection spread via contact interactions. Our simulations show that the number of successive waves of COVID-19 infection in specific countries depend on their underlying effective population density, the intermixing rate, and most importantly the timing and the duration of the control interventions imposed on/by the population. The effects of these interventions are visible as modulations in the infection rate.
As test cases, we compare our simulation results with reported data from India, Serbia, South Africa and USA. We show that multiple peak structures of the pandemic waves are reasonably well reproduced by the simulations. 
%
%
%

\section{The random walk Monte Carlo method}
Our random walk simulation model is described in our previous paper~\cite{st} and similar in approach to the work reported in~Ref.~\cite{Frasca}. Briefly, people belonging to a region of population $N$,  are described as points that execute random walks on a 2-dimensional plane. The speed of infection growth depends on the jump distance $l$ for each point, which we assume is a multiple of the mean separation $\langle r \rangle$ between points. In Cartesian coordinates, the jump components are simply $l \cos \theta$ and $l \sin \theta$, with $\theta$ generated randomly between $0$ and $2\pi$ from a uniform distribution.  
Unlike other approaches~\cite{Chu21,Kelker}, we do not consider a lattice or apply periodic boundary conditions. Instead, if a jump takes a point out of the area considered, it is reflected back into the system in a random direction.
In all cases we consider particles in a 1~km$^2$ unit area element. Therefore we can use the words population and population density interchangeably. For $N$ random walkers per unit area, $\langle r \rangle =\sqrt{1/N}$ in units of km. A separation distance of $\le 2$~m between individuals is taken as the `contact' distance~\cite{st}. 

The simulation starts with the introduction of an infected 
individual in state $I$. The disease then spreads through contact interactions, resulting in a drop in the number of susceptible points, $S$, which at time zero equals $N-1$.  Each time step corresponds to one random jump executed by all the points and we take that as one day. The number of infected points increase with such steps, based on the number of contacts between susceptible and infectious individuals. The populations in $S, I$ and $R$ states are $N_s,N_i$ and $N_r$ respectively, so that $N = N_s+N_i+N_r$ is preserved at all times. This results in an SIR compartmentalization of the population. 

In the next step of the simulations, a recovery rate was incorporated in a slightly different manner than used previously in Ref.~\cite{st}. 
%
We assume that 80\% of the infected population (picked at random) ultimately recover, becoming immune, and keep track of the number of time-steps (days) taken by an infected point before it goes into a recovered state. The time period is taken to be 35~days, and determined from a comparison of our simulated results with data. This choice of $\mu = 1/35~{\rm day}^{-1}$ is not unfounded. The incubation period of the coronavirus disease (after which symptoms develop) for infected individuals is found to be in the range of 8.2--15.6~days at the $97.5^{\rm th}$ percentile level~\cite{incub}. Furthermore, recent studies have shown that COVID-19 recovery times have an average value of about 25--28 days~\cite{Barman,Chae}. Given that these recovery times are evaluated after symptoms develop, a total of 35~days after infection is a reasonably accurate estimate.  

From here on, one can follow two equivalent approaches to incorporate the various intervention and mitigation strategies, usually employed after the start of an epidemic such as COVID-19.  A decisive criterion is the strict imposition of a lockdown that places significant mobility constraints on a majority of the population. This effectively moderates the growth in $N_i$, so that it is compensated by the recovery rate $\mu$. Other interventions such as vaccinations, mask mandates and behavioral changes by the population (social distancing, bubbles, etc.) additionally contribute to stalling the epidemic growth.  

In our earlier work~\cite{st}, we studied contact-interactions between random walkers on a plane, with $\beta$ equal to unity. This approach can be modified to incorporate the effects of all control interventions, exclusively via mobility restrictions on the stochastic agents. Such restrictions would impede the growth of the epidemic curve, during which time the number of recoveries continue to rise. However this is never a permanent solution. In more realistic scenarios the restrictions are relaxed from time to time (such as with lockdowns). In such a situation the epidemic growth continues, usually with a steep rise in the number of infections that is shifted in time.    

An alternative approach would be to assume $\beta$ values that are $< 1$ for specific cases under consideration, during the times that the control interventions and mitigation strategies are followed. For example, vaccinations, mask mandates, etc. lead to a drop in the number of susceptible individuals that can be infected. The effect of this drop can also be incorporated within the parameter $\beta$. This fractional $\beta$ is implemented by invoking a random number $r$ uniformly between 0 and 1, such that a change in state from $S \to I$ only occurs when $r \le \beta$. We show below that both approaches yield similar results with regard to predicting multiple infection waves. However we prefer the second approach, as the former may mislead the reader in terms of the difference between {\it actual} lockdowns imposed on a population and \textit{effective} mobility restrictions on the random walkers in the simulations.     

The simulations were performed for comparison with reported data 
for India, South Africa, Serbia and USA. The adjustable parameters were the population density (two values were used for this work, $N =$ 5k and 10k),  $\beta$, the percentage of mobile walkers, the duration during which mobility was restricted and the jump length $\l$. The recipe for the simulations was as follows. We consider the start date of the growth curve to be $D_0$. The intervention and mitigation strategies during this initial phase would effectively produce a first wave peak in $N_i$ on date $D_1$. This is obtained from reported data. The parameters in the simulation are then adjusted, so that the Monte Carlo results reproduce the observed first wave peak in the fractional daily infection rate.\footnote{We only consider events where the disease trajectory extends beyond day 500.} The simulated results are obtained from  $[N_r(t+\Delta t)-N_r(t)]/N$, for $\Delta t = 1$~day. Here, it should be noted that since $\partial N_r/\partial t \propto N_i$, the peaks in $N_i$ roughly coincide with the daily infection rate peaks along the time axis. These simulated results are compared to scaled-down data from the World Health Organization (WHO)~\cite{WHO}. This comparison is then used to infer if secondary or tertiary pandemic waves subsequently appear at later dates, due to the control interventions imposed in previous time windows. This is based on the premise that the control interventions and mitigation strategies would reduce the slope of the growth curve in the time duration that they are imposed (or followed).  At the end of each such duration (when restrictions, mandates etc. are relaxed) there would be growth again. This effectively results in growth curves with multiple peaks in daily infection rates, that are shifted in time and interpreted as epidemic waves. We discuss a few country specific results below.
 
\bef[htp]
\bc
 \includegraphics[scale=0.40]{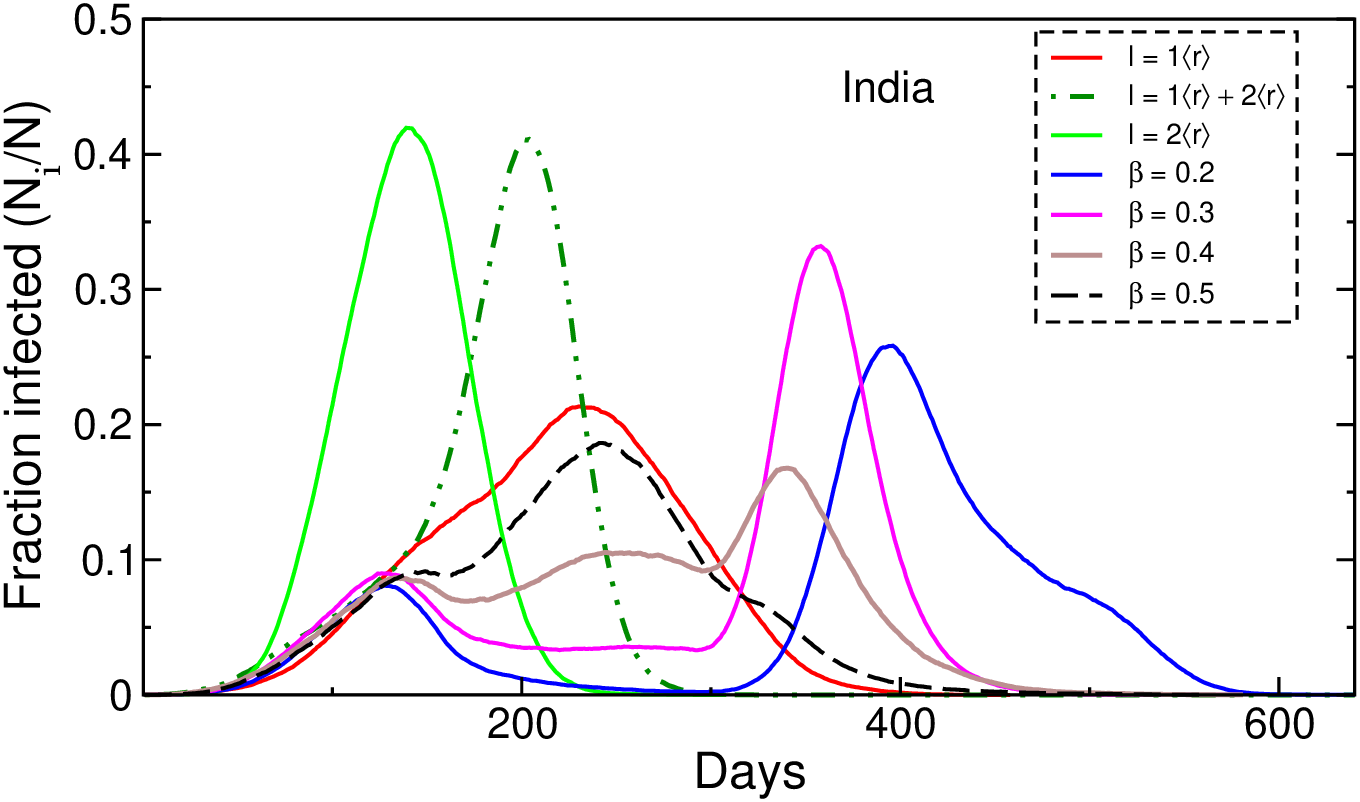}
 \caption{Simulated infected fractions for India, obtained for various parameter sets. The curves for whom the $\beta$ values are not specifed in the legend were generated assuming $\beta = 1$. The others were generated with the $l = 1 \langle r \rangle$ and $2 \langle r \rangle$ combination, as described in the text. All results are for $N = 10$k and averaged from five independent simulations.
}
 \label{fig1}  
\ec
\eef

\bef[htp]
\bc
 \includegraphics[scale=0.4]{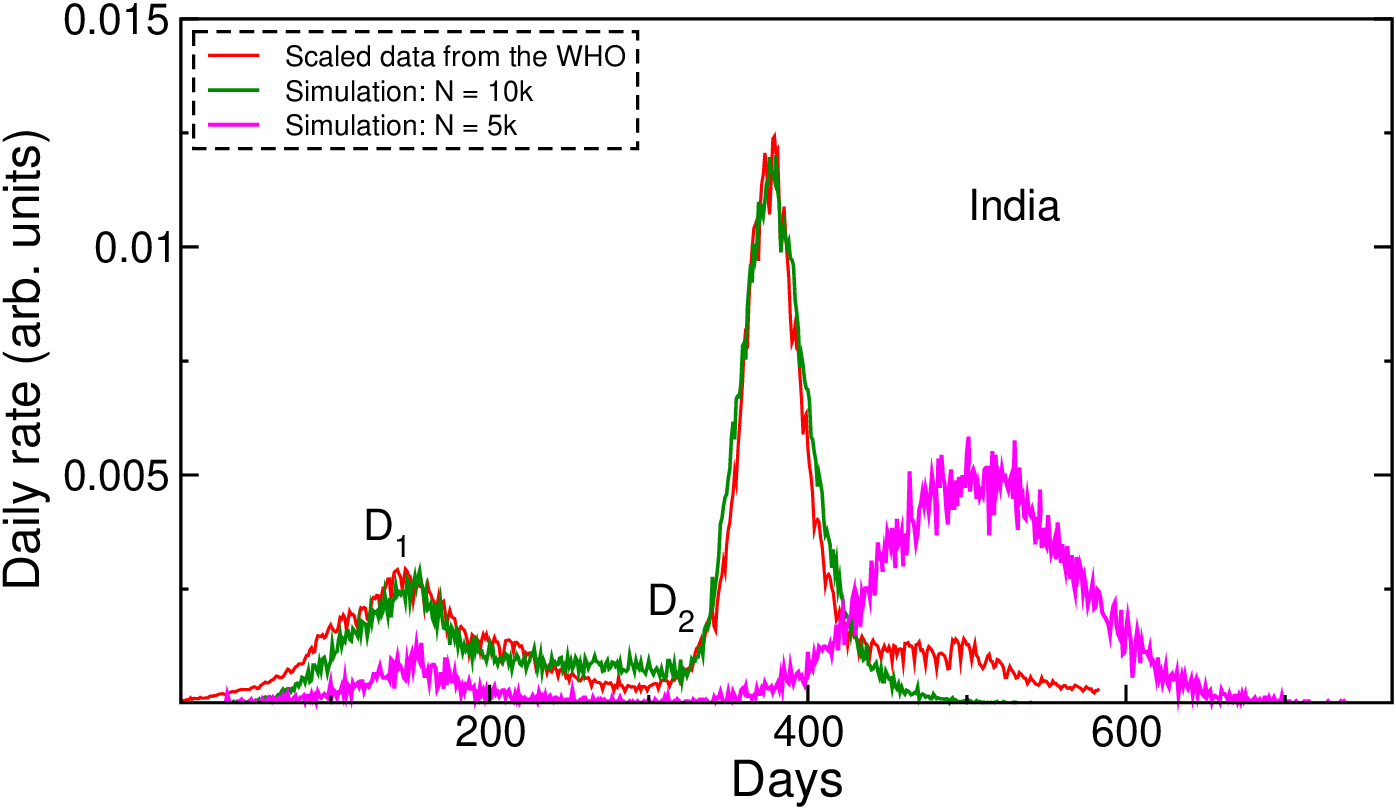}
 \caption  {Simulation results for India assuming two different densities ($N = 5$k and 10k), compared with normalized WHO data. The simulated values were averaged over six sets.}
 \label{fig2}  
\ec
\eef

\bef[htp]
\bc
\includegraphics[scale=0.4]{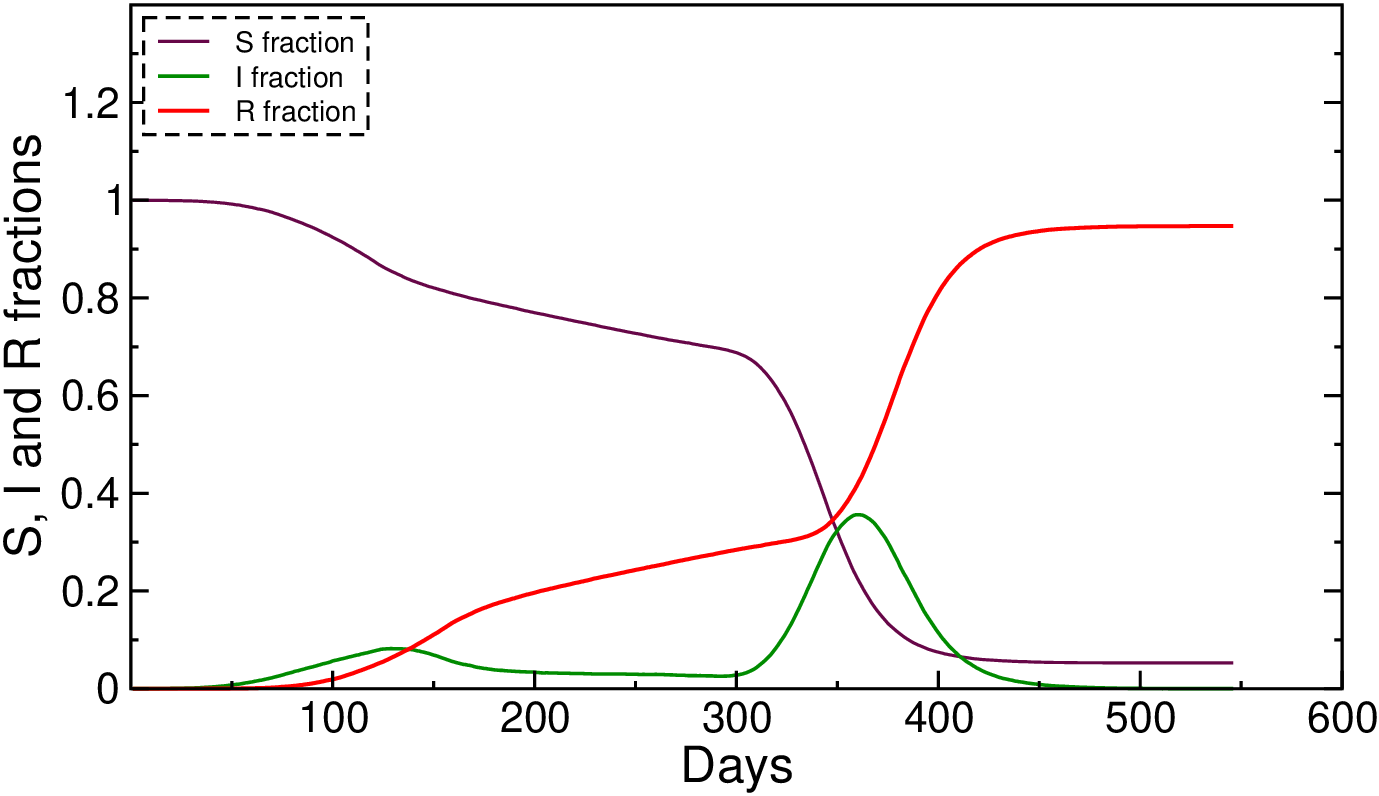}
 \caption  {SIR fractions for the $N = 10$k results in Fig.~\ref{fig2}.}
 \label{fig:SIR}  
\ec
\eef

\section{Analysis}
\label{sec:analysis}
\subsection{India}

We first consider the case of India, since the reported data show two distinct pandemic peaks. Furthermore, there has been a lot of speculation regarding the appearance of a large third wave, following the devastating second wave in 2021. The data show a first wave peak around Sep 15, 2020 and a second wave peak around May 06, 2021. These two structures contain in them all effects of 
interventions and mitigation strategies, including imposed lockdowns. We take $D_0$ to be April 22, 2020, with the first 
 peak ($D_1$) around day 150 (September 15, 2020). Since this was the first country that we studied, we performed several simulations to better understand the general shapes of the curves, obtained for different parameter values. The first two sets of simulations used jump steps of $l = 1\langle r \rangle$ and $2\langle r \rangle$, with $\beta = 1$. In the third set, we used $l = 1\langle r \rangle$ from $D_0$ to $D_1$ and $l = 2\langle r \rangle$ beyond $D_1$. The fractional $N_i$ results for different parameter values and $N = 10$k are shown in Fig.~\ref{fig1}. As evident in the figure, a larger amplitude infection peak is obtained when $l = 2\langle r \rangle$ step sizes are used, with a peak value of approximately 0.45 around day 140. For this case the number of infections drop to negligible levels after about 200~days. In comparison, $l=1\langle r \rangle$ produces a broader and smaller peak that is shifted to a later date (near day 230). On the other hand, the $l = 2\langle r \rangle$ and $l=1\langle r \rangle$ combination yields two distinct peak structures, as shown in the figure. In the next set of simulations we used this combination, together with $\beta$ values $< 1$ that account for intervention/mitigation strategies, which would lead to a drop in the $N_i$ fraction and consequently the daily infection rate. These results are also shown in Fig.~\ref{fig1}, for different values of $\beta$ during the time period $D_1$ to $D_2$ (the falling part of the first curve), with the later date representing the beginning of the second wave. Beyond $D_2$ we assume $\beta = 1$ for the rise of the second wave. Guided by reported data we take $D_2$ to be day 300. 
 
We observe that the shapes of the fractional $N_i$ curves, obtained with $\beta = 0.3$ from dates $D_1$ to $D_2$ (magenta curve in Fig.~\ref{fig1}) are very similar to scaled down daily infection rates obtained from the WHO, including the relative amplitudes for the two peaks. This is further validated in Fig.~\ref{fig2}, which shows a comparison between the reported daily rates and the simulated results (assuming $\beta=0.3$). 
The excellent agreement between the simulations and the reported data for the two waves is noteworthy. Similar agreement is not obtained from other parameter value combinations (such as the use of only $l = 1\langle r \rangle$ or a population density $N = 5$k for the $1\langle r \rangle + 2 \langle r \rangle$ jump-step combination. The results from the latter simulations are shown in Fig.~\ref{fig2} for comparison). The infection rate $\beta = 0.3$, used for time periods when effective mitigation/intervention measures are assumed in the simulations, is consistent with those obtained with a deterministic SIQR$_{\rm k}$ model that also took into account the quarantined and confirmed-recovered portions of the population (c.f. Table~2 in Ref.~\cite{Chae}.)

It is also important to point out here that in addition to the aforementioned agreement with data, the second wave peak in Fig.~\ref{fig2} emerges naturally from our simulations, as a result of interventions imposed previously, mainly through the parameter $\beta$. 
Furthermore, the simulations also show no indication of a third peak that may appear on applying additional interventions (through $\beta$) in the second wave. This is supported by Fig.~\ref{fig:SIR}, which shows the time development of the individual $S,I$ and $R$ components for these data. One can see that after day 500 $N_i$ drops to nearly zero, while $N_r$ saturates at $\sim 90\%$. It is apparent that at this point there are not enough infectious agents remaining to drive the curve and infect the small remainder of the susceptible population. This does not indicate a significant third wave. However, it may be noted that we assume all recovered individuals to be immune in our simulations. During the course of this work, several countries reported a surge in COVID-19 infections due to the emergence of the SARS-CoV-2 Omicron variant~\cite{omicron1}. This variant is characterized by an unusually large number of mutations in the spike protein and an ability to escape vaccine induced immunity~\cite{omicron2}. In light of this recent development, we performed
additional simulations that allow reinfections. Our preliminary results show that a small fraction of reinfections ($\lesssim 0.5\%$) would increase the susceptible population significantly, enough to cause a third wave for India.\footnote{For example, our preliminary simulations show that 0.2\% of reinfections would cause a third wave peak for India around day 600.} Further investigations that delve into this aspect and take into consideration the use of a different infection probability $\beta$ (due to the waning of vaccine-induced immunity) will be useful in this regard.
 
\subsection{Serbia}

Serbia is an interesting case study. The data show that the pandemic trajectory comprises two small peaks followed by two large peaks~\cite{WHO}.  The country was grappling with another wave of infections during the time of this work. It was 
therefore interesting to see whether the last wave is a consequence of the interventions imposed earlier. Here we take $D_0$ as March 04, 2020, following which there are four pandemic peaks in the daily rates (excluding the latest wave). In our simulations these correspond to time domains where $\beta$ values $< 1$ are used. We identify these to be days 50--100, 130--180, 250--320 and 380--450. Since Serbia has a much lower population density compared to India, we used $N = 5$k/unit area and $l = 2 \langle r \rangle$. Guided by the results for India, we again assumed $\beta = 0.3$ for the periods that correspond to systematic drops in the observed rates, following each observed peak in the epidemic curve. 
%

The simulation results are shown in Fig.~\ref{fig:Serbia}, together with corresponding scaled down data reported by the WHO. One can clearly see reasonable agreement between the two, with the simulations adequately predicting the peak of the fifth wave that was ongoing at the time of this work.

\bef[htp]
\bc
 \includegraphics[scale=0.4]{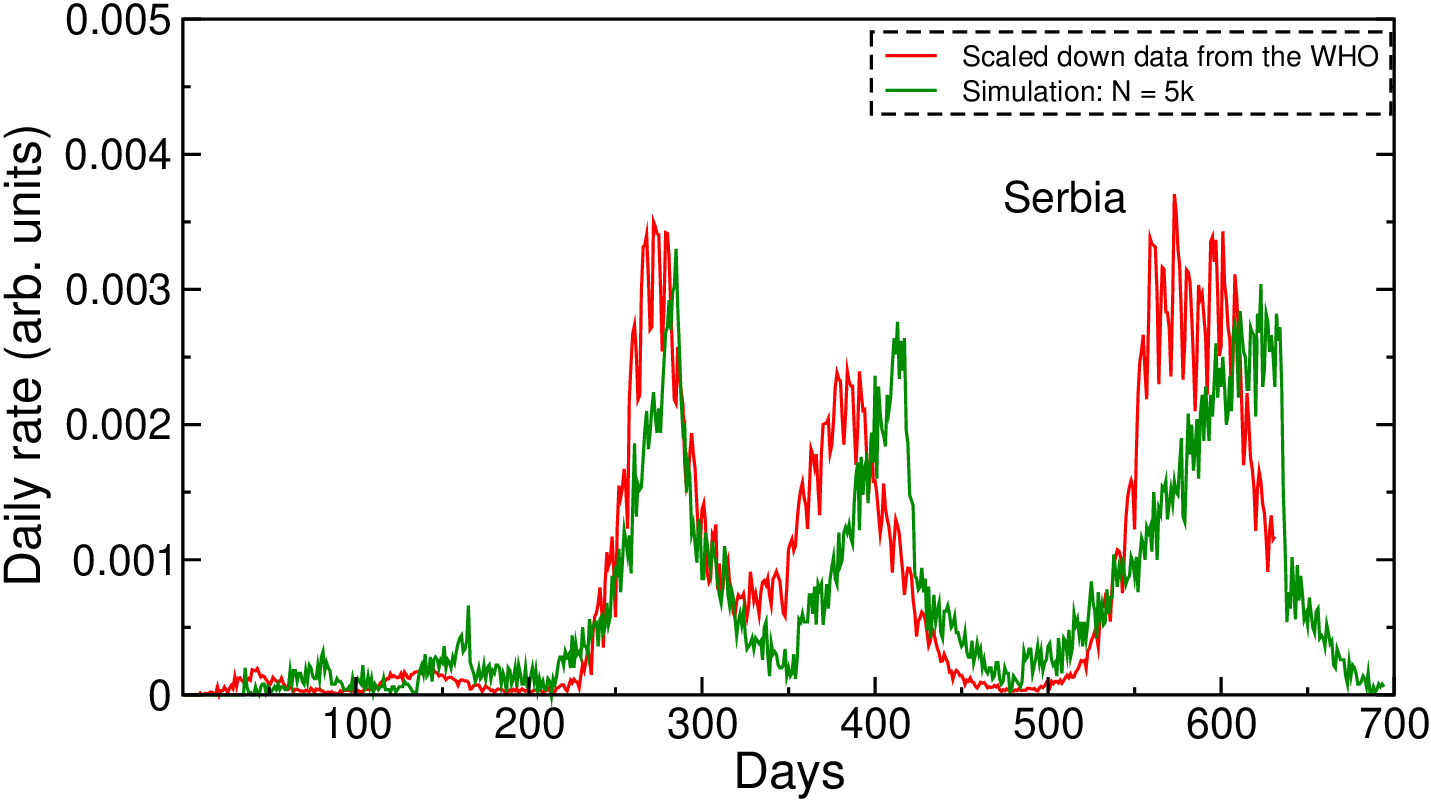}
 \caption  {Simulated daily rates of Serbia, compared 
 with scaled down WHO data. The simulation results were averaged over 10 sets.}
 \label{fig:Serbia}  
\ec
\eef

\bef[htp]
\bc
 \includegraphics[scale=0.4]{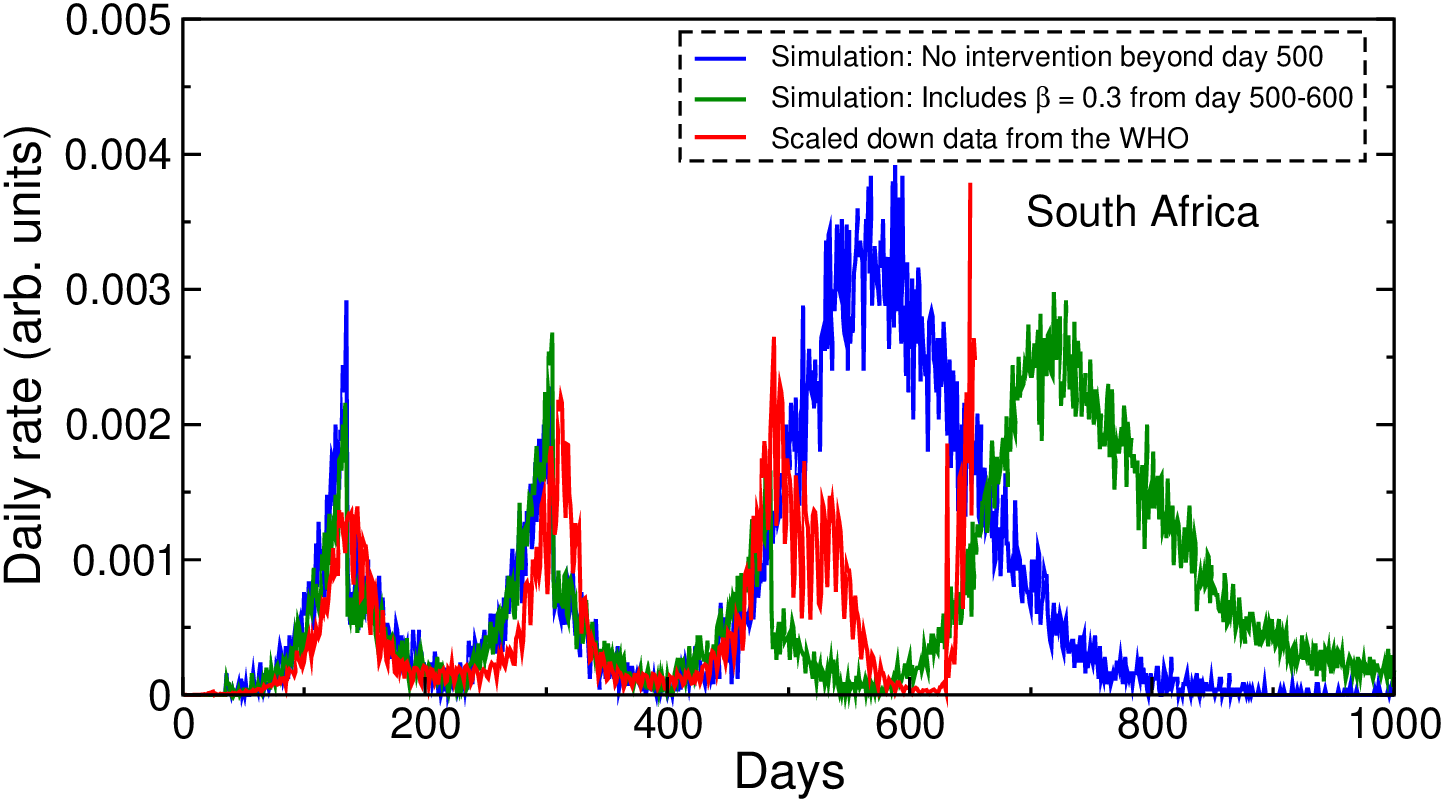}
 \caption  {Simulated results for South Africa, compared 
 with scaled down WHO data. The simulations assumed $N = 5$k and were averaged over 10 sets. The blue curve is generated assuming no control interventions beyond day 500. The green curve assumes such interventions ($\beta = 0.3$) from day 500--600.}
 \label{fig:SA}  
\ec
\eef

\subsection{South Africa}
Next we looked at the data for South Africa, whose growth trajectory is interesting to follow, as the Omicron variant was first identified in the region~\cite{omicron1}. 
We took $D_0$ to be Mar 04, 2020. There are three 
pandemic waves, with peaks around days 150, 300 and 500. To simulate the initial wave we used\footnote{South Africa also has a lower population density, similar to Serbia.} $N = 5$k and effective interventions ($\beta = 0.3$) from day 150 to 250, with $l = 2 \langle r \rangle $. On imposing a second period of lower infections from day 320 to day 420, we obtain consistent results for the first two waves. These are plotted in Fig.~\ref{fig:SA}. 
In comparison, the peak around day 500 was found to be too broad compared to the reported data from the WHO. We note that a further intervention period from day 500 to day 600 (again with $\beta = 0.3$) yields daily rates for the third wave that are in reasonably good agreement with WHO data. These results also show a clear fourth wave, whose position depends on the imposed intervention in the previous wave. This Omicron driven fourth COVID-19 wave has already been reported for South Africa~\cite{fourth}. It is interesting to note that our predicted fourth wave is rather broad, similar to the blue curve obtained on assuming $\beta = 1$ beyond day 500. Although we have not carried out simulations with imposed control interventions beyond day 700, this similarity suggests the possibility of a fifth wave for South Africa.

%

\bef[htp]
\bc
 \includegraphics[scale=0.4]{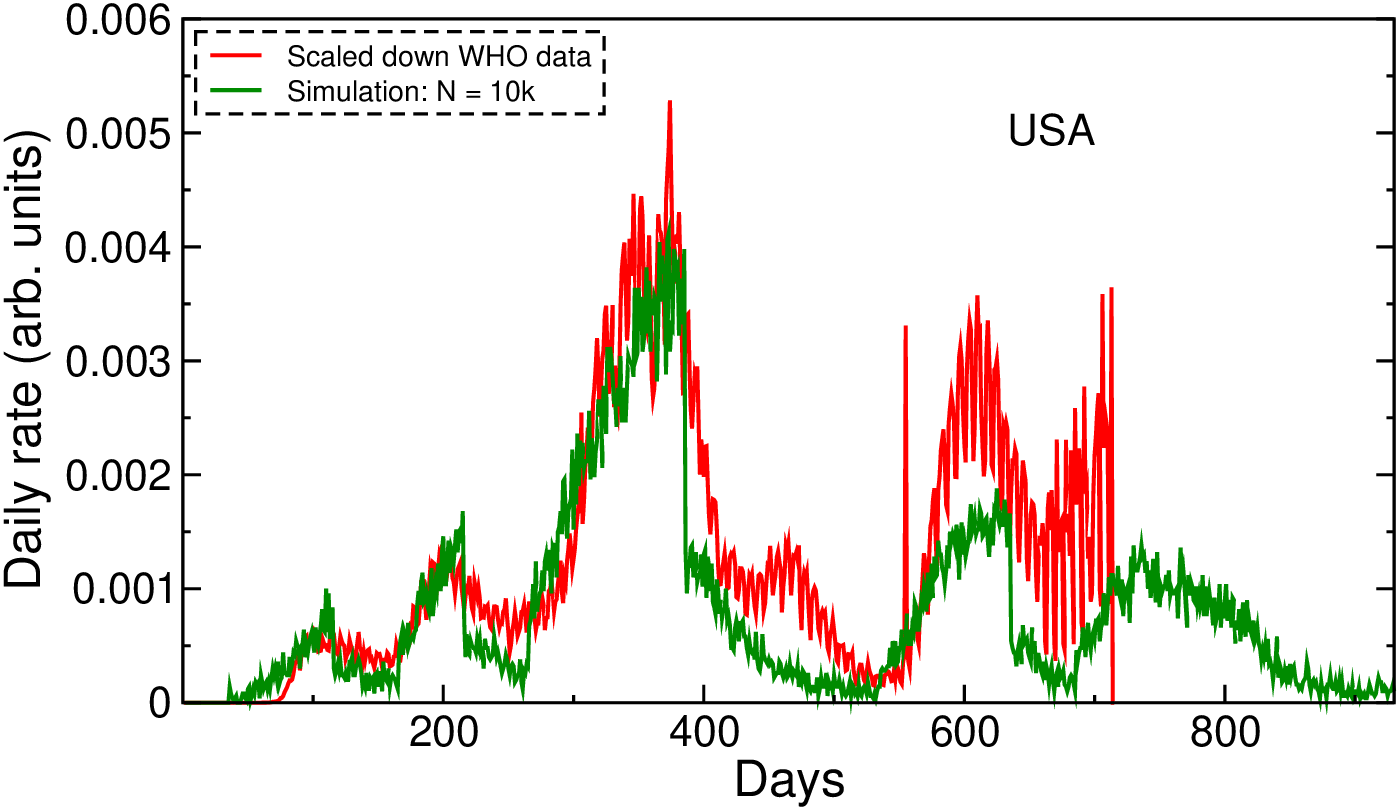}
 \caption  {Simulated rates for USA compared with scaled-down WHO data. These  simulations assumed $N = 10{\rm k}, l = 0.7 \langle r \rangle$ and were averaged over 5 runs.}
 \label{fig:USA}  
\ec
\eef

\subsection{USA}
Our final set of simulations were for USA, which presents a different challenge as each of its 50 states follow independent mitigation and containment strategies. There are five observed waves of infections, with the third peak being significantly larger than the others. A scaled down version of the reported data is shown in Fig.~\ref{fig:USA}. In our simulations we use $N = 10$k and take $D_0$ to be January 1, 2020. Based on the WHO data, we introduce three intervention periods, between days 130--180, 230--280, and 400--550. For these time periods we used $l = 0.7\langle r \rangle$ and $\beta = 0.3$. As Fig.~\ref{fig:USA} shows, apart from the small peak around day 470,  all other features in the data are reasonably well reproduced by the simulations. The simulations also correctly predict a wave after day 500, which was ongoing during the time of this work. On the basis of our results one can also anticipate a fresh wave of infections following the peak after day 600. Similar to the other cases studied, we observe that control interventions following an increase in the daily rate of infections invariably lead to a new epidemic wave later in time, provided there are enough agents available to infect the remaining susceptible fraction.

\section{The equivalence between mobility restrictions on the random walkers and imposition of a lower infection rate}

One can get similar results as above by using the other approach that assumes $\beta = 1$ at all times. This is obtained by applying mobility restrictions at different time ranges on most of the random walkers, so that the primary infection peaks are reproduced.  However, it is important to keep in mind that these time ranges are not the same as periods of imposed lockdowns by governing bodies.
We discuss below the results obtained using this analogous approach, for the case of India. 

Here again two different sample populations with $N = 5$k and 10k were used to simulate the data for different jump lengths $l$. All calculations were carried out for $\mu = 1/35$~day$^{-1}$ with a 20\% mobile population during the time period spanning from day 120 to day 300. The remaining 80\% remain frozen at their positions.
Fig.~\ref{fig:1ld2r_data} shows the fractional daily rates obtained for the two populations for the intervention period mentioned above, together with a jump distance of $l = 2\langle r \rangle$.  The 10k results show excellent agreement with reported data for the second wave. Similarly consistent results are obtained for the other countries studied in this work. 

\bef[t]
\bc
 \includegraphics[scale=0.4]{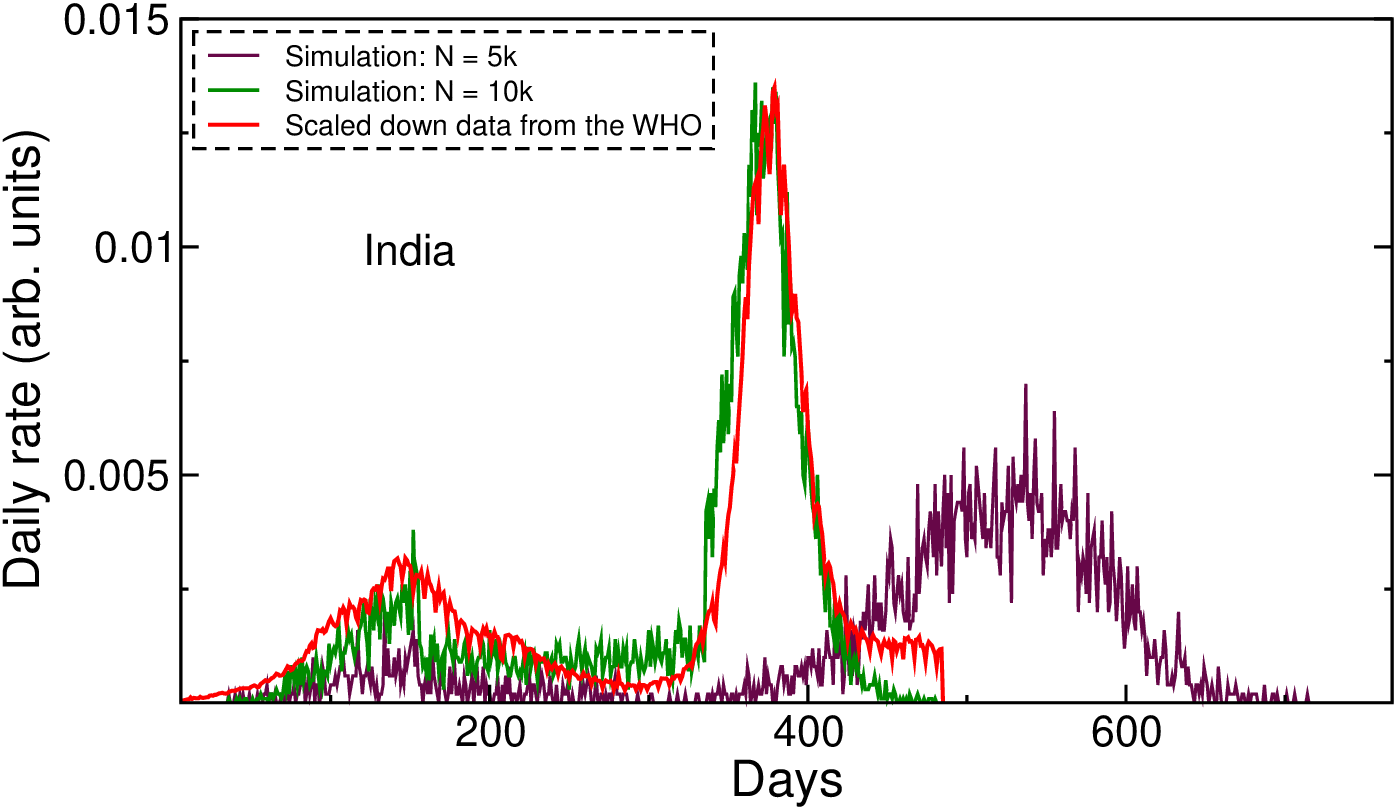}
 \caption  {Simulated daily rates for India obtained using the mobility restrictions as described in the text and $l = 2 \langle r \rangle$. The $N = 10$k results are averaged over 7 sets, while the $N = 5k$ results are from a single set, shown for comparison with normalized WHO data.}
 \label{fig:1ld2r_data}  
\ec
\eef


\section{Discussion}

The results presented in this work show that it is possible to predict the  trajectory of an epidemic using a random walk Monte Carlo approach that includes time periods of control interventions. Two equivalent simulation approaches (either using $\beta < 1$ or keeping a percentage of the population immobile when effective interventions were in place) can be followed, as they yield similar results. We observe that for populous countries such as India and USA
reasonably good results are obtained on using a higher population density (10k) in the simulations.\footnote{Although not shown explicitly, we find that a higher population of $20$k per unit area leads to poor results.} 
In comparison, for countries such as Serbia and South Africa (whose population densities are much smaller) a $N = 5$k works best for the simulations. Conservatively, both these numbers take into account higher floating densities at localized places. In all cases, we show that the use of step-sizes $\lesssim 2\langle r \rangle$ yield optimal results. The bound on this parameter value can be justified as below. 

In our earlier work~\cite{st} we showed that there are essentially two limits of growth. If the jump step is too small then the growth is purely quadratic in time~\cite{axel} and if the jump step is too large the
growth is exponential due to homogeneous
mixing~\cite{Buscarino}. The country-specific growth modes described here occur between these two limits when there is sub-exponential behavior~\cite{maier} due to the containment strategies employed. The power-law nature of the growth is lost for jump steps $\ge 3 \langle r \rangle$. On the other hand when the density becomes
too high, since our simulations use jump lengths that are proportional to the average inter-person separation, the epidemic growth approaches the quadratic (lower) limit. In such a scenario an alternative approach, such as in Ref.~\cite{Sausa} may be better suited for studying the sub-features in a
single wave. This may offer an explanation as to why our simulated results for large jump lengths or population densities fail to reproduce the observed data.  

It must also be added that in the present formalism 
the exact total number of cases are not meaningful. Instead, the focus is on the rapid time-shifted increase in the number of infections following intervention periods. These correspond to new waves of 
infection. On this basis the simulations are used to predict pandemic peaks at  later dates, based on earlier data. If one chooses to use a calibration factor based on previously observed peaks, it is also possible to roughly estimate the total number of people infected.

\section{Conclusions}

To conclude, we use a random walk Monte Carlo simulation technique to better understand contact-based epidemic spread. The model of independent random walkers as infection carriers in two-dimensional space is intuitive and shows promise in better understanding infectious disease outbreaks~\cite{st}. It is more accessible than other models and has the ability to capture random interactions that are missed through more conventional approaches. The simulations are shown to predict reasonably accurate disease trajectories in terms of secondary and tertiary waves, for COVID-19 data from four countries with vastly different features. 
\section{Acknowledgements}
We are grateful to Prof. Niranjan Barik for several helpful discussions and to Dr.~Ara Chutjian for useful remarks. ST acknowledges funding support from the National Research Foundation (South Africa) under grant number 85100.  
\bibliography{ld_sim_resubmit}

\end{document}